\documentclass[a4paper,11pt]{article}
\pdfoutput=1 

\usepackage{jheppub} 

\usepackage[T1]{fontenc} 
\usepackage{arydshln}
\usepackage{slashed}
\usepackage{multirow}

\def\a{\alpha}
\def\b{\beta}

\def\D{\Delta}

\def\e{\epsilon}

\def\m{\mu}
\def\n{\nu}

\def\ps{\psi}
\def\ph{\phi}
\def\Ph{\Phi}

\def\t{\tau}

\def\ch{\chi}

\def\pd{\partial}

\def\scO{\mathcal{O}}

\def\scD{\mathcal{D}}

\makeatletter
\newsavebox\myboxA
\newsavebox\myboxB
\newlength\mylenA
\newcommand*\xoverline[2][0.75]{%
    \sbox{\myboxA}{$\m@th#2$}%
    \setbox\myboxB\null
    \ht\myboxB=\ht\myboxA%
    \dp\myboxB=\dp\myboxA%
    \wd\myboxB=#1\wd\myboxA
    \sbox\myboxB{$\m@th\overline{\copy\myboxB}$}
    \setlength\mylenA{\the\wd\myboxA}
    \addtolength\mylenA{-\the\wd\myboxB}%
    \ifdim\wd\myboxB<\wd\myboxA%
       \rlap{\hskip 0.5\mylenA\usebox\myboxB}{\usebox\myboxA}%
    \else
        \hskip -0.5\mylenA\rlap{\usebox\myboxA}{\hskip 0.5\mylenA\usebox\myboxB}%
    \fi}
\makeatother

\newcommand{\be}{\begin{eqnarray}}
\newcommand{\ee}{\end{eqnarray}}
\def\a{\alpha}
\def\b{\beta}

\title{2, 84, 30, 993, 560, 15456, 11962, 261485,$\ldots$: Higher dimension operators in the SM EFT}

\author[a]{Brian Henning,}
\author[b]{Xiaochuan Lu,}
\author[c,d]{Tom Melia}
\author[c,d,e]{and Hitoshi Murayama}

\affiliation[a]{Department of Physics, Yale University, New Haven, Connecticut 06511, USA}
\affiliation[b]{Department of Physics, University of California, Davis, California 95616, USA}
\affiliation[c]{Department of Physics, University of California, Berkeley, California 94720, USA}
\affiliation[d]{Theoretical Physics Group, Lawrence Berkeley National Laboratory, Berkeley, California 94720, USA}
\affiliation[e]{Kavli Institute for the Physics and Mathematics of the Universe (WPI), Todai Institutes for Advanced Study, University of Tokyo, Kashiwa 277-8583, Japan}

\emailAdd{brian.henning@yale.edu}
\emailAdd{xclu@ucdavis.edu}
\emailAdd{tmelia@lbl.gov}
\emailAdd{hitoshi@berkeley.edu, hitoshi.murayama@ipmu.jp}

\preprint{IPMU15-0207}

\abstract{In a companion paper \cite{Henning:2017fpj}, we show that operator bases for general effective field theories are controlled by the conformal algebra. Equations of motion and integration by parts identities can be systematically treated by organizing operators into irreducible representations of the conformal group. In the present work, we use this result to study the standard model effective field theory (SM EFT), determining the content and number of higher dimension operators up to dimension 12, for an arbitrary number of fermion generations. We find additional operators to those that have appeared in the literature at dimension 7 (specifically in the case of more than one fermion generation) and at dimension 8.
\newline
\newline

{\small
\noindent (The title sequence is the total number of independent operators in the SM EFT with one fermion generation, including hermitian conjugates, ordered in mass dimension, starting at dimension 5. )
}
}

\notoc 
\begin{document} 
\maketitle
\flushbottom
\newpage

\section{Introduction}
\label{sec:intro}

In this paper we apply a newly developed technique which counts the dimension of operator bases in effective field theories (EFTs) to the standard model (SM) EFT. We demonstrate counting up to dimension 15, allowing for an arbitrary number of fermion generations $N_f$, and provide explicit results for the content of the independent operators up to dimension 12. 

Previous results have appeared in the literature at dimension
 5 \cite{Weinberg:1980bf}, 6 \cite{Buchmuller:1985jz,Grzadkowski:2010es,Alonso:2013hga}, 7 \cite{Lehman:2014jma} and very recently 8 \cite{Lehman:2015coa}. This particular counting problem has a history of being tricky, due to the somewhat complicated nature of the particle content of the SM, and the intricacies of equations of motion (EOM) and integration by parts (IBP), which yield relations between operators. With our new method, the procedure is for the first time completely automated, reducing the possibility for error by putting EOM and IBP on the same footing as gauge and global symmetries, which can then be systematically dealt with by using group theoretic techniques. Enumeration of operators is encoded in a Hilbert series, which we compute by making use of the plethystic exponential and Molien's formula, as reviewed \textit{e.g.} in the phenomenological papers \cite{Jenkins:2009dy, Hanany:2010vu, Lehman:2015via}.
 
There are two key observations in our technique, both of which rely on the conformal group.  (1) The EOM can be regarded as an ideal in a commutative ring, so that it can be implemented in the Hilbert series.  Accounting for EOM leads to generators which fall into short multiplets of the conformal group.  (2) The IBP removes operators that are total derivatives, which can be regarded as descendants in a conformal field theory.  Therefore, we only need to identify the primary fields.  Note that we regard the SM with only the kinetic terms as the zeroth order Lagrangian, where all fields are massless and hence the theory is (classically) conformal.  The we classify additional higher-dimension operators as perturbation to the system.  This technique is a natural generalization of what we studied in 1D QFT in our previous paper~\cite{Henning:2015daa}.

In what we intend to be the final twist in the tale of the SM EFT operator basis dimension, we find a correction to the $N_f>1$ counting at dimension 7 and an additional 62 operators at dimension 8 to those previously found in the literature at $N_f=1$. The method is strikingly simple, as exemplified in the accompanying \texttt{Mathematica} notebook, which we encourage the interested reader to study. In this paper we aim to present the method with a minimal amount of technical details, but to the level at which it can be reproduced and applied to other phenomenological Lagrangians of interest.

While the SM EFT is of obvious phenomenological utility, study of it has also led to interesting theoretical questions and developments. Our investigations are motivated by the difficulties in determining the operator content of the SM EFT, as well as interesting features so far uncovered in the well developed literature~\cite{Buchmuller:1985jz,Grzadkowski:2010es,Alonso:2013hga,Lehman:2014jma,Lehman:2015coa,Lehman:2015via,Grojean:2013kd,Jenkins:2013zja,Jenkins:2013wua,Elias-Miro:2013gya,Elias-Miro:2013mua,Elias-Miro:2013eta,Manohar:2013rga,Hartmann:2015oia,Alonso:2014zka,Jenkins:2013fya,Jenkins:2013sda,Alonso:2014rga,Elias-Miro:2014eia,Cheung:2015aba,Gupta:2014rxa,Trott:2014dma,Henning:2014wua,Drozd:2015rsp,Drozd:2015kva,Huo:2015nka,Gorbahn:2015gxa,Ghezzi:2015vva,Masso:2014xra,Ellis:2014jta,Falkowski:2014tna}.

The method is outlined in section~\ref{sec:method}; details beyond those needed for the present purpose can be found in \cite{Henning:2017fpj}. The ingredients to apply this method to the SM EFT are explicitly laid out in section~\ref{sec:HSSM}. We present some selected results in section~\ref{sec:results}: we give corrected versions of the counting of operators at dimension 7 and at dimension 8 for general $N_f$; we give the number of independent operators as a function of $N_f$ and split according to baryon violating number, $\Delta B$, up to dimension 12, and up to dimension 15 retaining only $N_f$ dependence. Hilbert series up to dimension 12 can be found in an accompanying \texttt{Mathematica} notebook. Section~\ref{sec:discuss} provides a short discussion on the application to other phenomenological Lagrangians. 

\section{Method}
\label{sec:method}

An operator basis for an EFT is a set of local operators which give independent \(S\)-matrix contributions. Redundancies associated with IBP and EOM must be accounted for in the operator basis. It turns out that the conformal group organizes an operator basis, allowing us to systematically handle IBP and EOM redundancies. A thorough investigation of this structure and its implications is carried out in~\cite{Henning:2017fpj}; here we provide only an outline of the method to understand what is necessary for application of the final formula, eq.~\eqref{eq:HSSM}, to the SM EFT. 

The emergence of the conformal group in this problem and the main result of~\cite{Henning:2017fpj} are schematically easy to understand. In its region of validity, an EFT is perturbative around a free field theory. Free field theories are conformal, hence the operator-state correspondence tells us that the space of local operators modulo EOM fall into representations of the conformal group. Representations of the conformal group consist of a primary operator \(\scO\) together with an infinite tower of descendant operators obtained by repeated differentiation of \(\scO\), \textit{i.e.} \((\scO,\pd_{\m}\scO,\pd_{\m_1}\pd_{\m_2}\scO,\dots)\). Importantly for our purposes, descendant operators are total derivatives.

All local operators can be constructed as products of some generating set of operators; these generating operators fill out representations of the conformal group themselves, so it is natural to form the space of local operators via tensor products of the generating representations. By decomposing these tensor products back into representations of the conformal group, IBP relations can be completely accounted for. In particular, only primary operators are not total derivatives. For a primary operator to be a part of the operator basis, it also needs to be a Lorentz scalar. We conclude that the operator basis consists of the scalar (spin 0) conformal primaries in the corresponding free field theory~\cite{Henning:2017fpj}.

As a way to study the operator basis, we wish to compute an object called the Hilbert series, which abstractly can be viewed as a partition function on the operator basis. More concretely, it is a generating function defined to count the number of operators of some given field content,
\begin{equation}
H(\scD,\{\ph_a\}) = \sum_{r_1,\dots,r_N}\sum_{k} c_{\mathbf{r}\, k} \ \ph_1^{r_1}\dots \ph_N^{r_N} \scD^k,
\label{eq:hil_def}
\end{equation}
where \(c_{\mathbf{r}\, k} \in \mathbb{N}\) is the number of independent, Lorentz invariant, gauge invariant operators composed of \(\mathbf{r} = (r_1,\dots,r_N)\) powers of \(\ph_1,\dots,\ph_N\) and \(k\) derivatives. In the above, \(\{\ph_a\}\) are ``spurions'' to label the content of operators; we stress that they are not fields, just complex numbers. Similarly, \(\scD\) is a spurion to represent the covariant derivative \(D\).

The method to compute the \(c_{\mathbf{r}\, k}\) builds upon our earlier work in $(0+1)$ dimensions \cite{Henning:2015daa}, where we used an underlying $SL(2,\mathbb{C})$ structure to organize operators into irreducible representations of the group; the operator basis was spanned by operators of highest weight of the $SL(2,\mathbb{C})$ representations. This is because in each irreducible representation, operators obtained from the one of highest weight by the lowering operators are total derivatives (the lowering operator is the derivative). A weighted sum and integral over the maximal compact subgroup of $SL(2,\mathbb{C})$, namely \(SU(2)\), was used to project out
the highest weight operators.

The analogous situation in $d$ dimensions is that $SL(2,\mathbb{C})\simeq SO(1,2)$ is replaced by the conformal group in $d$ dimensions, $SO(d+2, \mathbb{C}) \simeq SO(d,2)$. As described above, the operator basis is spanned by the set of scalar conformal primaries formed from the fields and their covariant derivatives. Echoing the one dimension case, the primary operators are projected out and counted via a weighted sum and integral over \(SO(d+2)\), the maximal compact subgroup of \(SO(d+2,\mathbb{C})\), of a generating function for tensor products of the generating representations.

Let us briefly describe these so-called generating representations. They consist of the fundamental fields \(\Ph_a\) (in four dimensions, scalars \(\ph\), fermions \(\ps\), or self-dual/anti-self-dual field strengths \(X_{L/R} = X \pm i \tilde{X}\)) together with an infinite tower of symmeterized derivatives\footnote{When there is no gauge symmetry, \(D_{\m} = \pd_{\m}\) and the derivatives are automatically symmetric. With gauge symmetries, the anti-symmetric combination is proportional to field strengths, \([D_{\m},D_{\n}] \sim X_{\m\n}\), and therefore already accounted for.} acting on \(\Ph_a\) with the EOM removed. Removing terms proportional to the EOM in this tower means eliminating terms proportional to \(D^2\ph\) for scalars, \(\slashed{D}\ps\) for fermions, or \(D_{\m}X_{L/R}^{\m\n}\) for field strengths. This construction is also described by Lehman and Martin~\cite{Lehman:2015coa}. In the present work, we make the observation that the tower of \(\Ph_a\) and derivatives acting on \(\Ph_a\) with EOM removed falls into an irreducible representation of the conformal group. In particular, these are short representations of the conformal group that correspond to states saturating a unitarity bound (the EOM acts as a shortening condition for the multiplet).

\subsection{Computing the Hilbert series}
The computation of the Hilbert series proceeds via elementary use of characters and representation theory. Here we only provide a schematic picture of the procedure and then quote the final formula, eq.~\eqref{eq:HSSM}, that is our launching point.

We start with the character for each generating representation, which is a product of its character under the conformal group and its character under the gauge symmetries,
\begin{equation}
\ch_{\ph_a} = \ch_{\ph_{a},SO(d+2,\mathbb{C})}\ch_{\ph_a,\text{gauge}}.
\label{eq:char_field}
\end{equation}
We recall that the character for a representation \(R\) of a Lie group \(G\) is given by \(\ch^{}_R(g) = \text{Tr}_R(g)\) for \(g \in G\). For connected \(G\), all \(g \in G\) may be conjugated into a maximal torus \(U(1)^r \subset G\) with \(r = \text{rank}\, G\). Since the character is a class function, \textit{i.e.} it is conjugation invariant since \(\text{Tr}(hg h^{-1}) = \text{Tr}(g)\), it is a function only of the \(r\) parameters \((x_1,\dots,x_r)\) of the torus which we denote by writing \(\ch^{}_R(x_1,\dots,x_r)\). Characters of the conformal group in arbitrary dimension are discussed in~\cite{Dolan:2005wy}, as well as in four dimensions in~\cite{Barabanschikov:2005ri}; their construction is also reviewed in~\cite{Henning:2017fpj}.

Taking tensor products of the generating representations amounts to character multiplication. A generating function known as the plethystic exponential is used to form all possible tensor products. The plethystic exponential accounts for the statistics of the underlying fields (symmetric for bosons, anti-symmetric for fermions); this is, more or less, the meaning of ``plethysm''.\footnote{The words Hilbert series, Molien's formula, and plethystic exponential often appear together. They all have roots in the subject of invariant theory and are closely related~\cite{Sturmfels:inv}. In the context of invariant theory, the Hilbert series is a partition function for a ring of invariants, Molien's formula is a way to compute the Hilbert series via a taking a tensor product of representations and averaging over the group, and the plethystic exponential is the integrand of Molien's formula.} For a spurion \(\ph_R\) in representation $R$, the plethystic exponential is defined as
\be
\text{PE}\big[\ph^{}_R\ch^{}_R(x_1,\dots,x_r)\big] =\exp\bigg(\sum_{n=1}^\infty \, \frac{1}{n} \,\,(\pm1)^{n+1} \ph_R^n \,\, \chi^{}_R(x_1^n,\dots,x_r^n)\bigg) 
\ee
where the $+(-)$ sign is taken for spurions associated with bosons (fermions). The formula may be more illuminating by recognizing that it simply comes from using \(\log\det = \text{Tr}\log\) and expanding a logarithm,
\begin{align}
\text{bosons: }& \frac{1}{\text{det}_R(1- \ph_R g)} = \exp\bigg(\sum_{n=1}^{\infty}\, \frac{1}{n} \ph_R^n\, \text{Tr}_R(g^n) \bigg) \, ,\\ 
\text{fermions: }& \text{det}_R(1+\ph_R g) = \exp\bigg(\sum_{n=1}^{\infty}\, \frac{1}{n}\,(-1)^{n+1} \ph_R^n\, \text{Tr}_R(g^n) \bigg) \, .
\end{align}
These equations also make it obvious that the plethystic exponential obeys
\begin{equation}
\text{PE}\big[\ph^{}_R\ch^{}_R\big]\text{PE}\big[\ph^{}_{R'}\ch^{}_{R'}\big] = \text{PE}\big[\ph^{}_R\ch^{}_R+ \ph^{}_{R'}\ch^{}_{R'}\big],
\end{equation}
with appropriate care taken for bosonic or fermionic statistics.

The Hilbert series is obtained by counting all the gauge invariant, scalar primaries that show up in tensor products from the plethystic exponential, weighted as in eq.~\eqref{eq:hil_def}. Scalar conformal primaries are projected out via character orthogonality by multiplying by all possible \(\ch^*_{SO(d+2,\mathbb{C}),\text{scalar}}\) and integrating over the conformal group. Gauge singlets are projected out by integrating over the gauge group.\footnote{This is also character orthogonality, where we multiply by the character of the singlet representation, which is unity, and integrate over the group.}

Following this procedure, we arrive at a formula for the Hilbert series given by~\cite{Henning:2017fpj}
\be
H(\scD,\{ \phi_a\}) = \int d\mu_{\text{Lorentz}}\int  d \mu_{\text{gauge}} \,\,\frac{1}{P} \,\, \text{PE}\Big[\sum_a \frac{\ph_a}{\scD^{d_a}}\ch_a\Big] +\Delta H(\scD,\{ \phi_a\}) \,\,.
\label{eq:HSSM}
\ee
This formula generates the Hilbert series which counts Lorentz and gauge singlets modulo EOM and IBP. Here \(d_a\) is the canonical mass dimension of the field \(\ph_a\) and the integrals are over Lorentz and gauge group parameters, with group measures $d\mu$. The $\Delta H$  is a small modification term due to subtle issues regarding lack of orthonormality of conformal characters (arising because the conformal group is non-compact), whose form is given for a general EFT in~\cite{Henning:2017fpj}. We evaluate $\Delta H$ explicitly for the SM below---importantly, it is comprised only of terms with mass dimension four and less.  To arrive at the first term on the rhs of eq.~\eqref{eq:HSSM}, we performed the integral associated with dilatations in the conformal group, leaving a remaining integral over the Lorentz group (this step is the generalization of performing the \(\a\) integral in~\cite{Henning:2015daa}). The factor
\begin{equation}
\frac{1}{P} \equiv \text{det}_{\Box}(1- \scD g),
\label{eq:1overP_def}
\end{equation}
where the determinant is taken over the vector \((\Box)\) representation of \(g \in SO(d)\), is a remnant of the Haar measure for the conformal group. The function \(P\) plays an important role in the connection with characters of the conformal group, as it is the generating function for symmetric products of the vector representation (recall that a conformal representation contains an infinite tower of symmeterized derivatives acting on a primary operator), see eq.~\eqref{eq:char}.

\subsection{A viewpoint of IBP redundancy using differential forms}
\label{subsec:HodgeDual}

The IBP redundancy states that total derivative operators are zero. In our method, this redundancy is accounted for by throwing away all the descendants while only keeping the primaries in each irreducible conformal group representation. However, with a little help from the Hodge dual of differential forms, one can obtain a quite useful alternative treatment of the IBP redundancy, which reveals the cohomological nature of this problem. This picture makes contact with the way that IBP relations were discussed for one spacetime dimension in~\cite{Henning:2015daa}, where it is explained how IBP relations for operators composed of \(r_1,\dots,r_N\) powers of \(\ph_1,\dots,\ph_N\) and \(k\) derivatives (schematically \(\ph_1^{r_1}\dots\ph_N^{r_N}\pd^k\)) arise from considering a total derivative acting on the independent operators with one less derivative (schematically \(0 = \pd(\ph_1^{r_1}\dots\ph_N^{r_N}\pd^{k-1})\)). We will see that the language of differential forms allows a straightforward answer to this question. 

Since we are counting Lorentz scalar operators, any total derivative scalar operator must have the form $\partial_\alpha A_\alpha$, where $A_\alpha$ is a Lorentz $1$-form operator. In $d$ dimensions, the Hodge dual of each scalar operator is a $d$-form; the dual of each total derivative operator $\partial_\alpha A_\alpha$ is an exact $d$-form:
\begin{align}
*\left( {{\partial _\alpha }{A_\alpha }} \right) &= \left( {{\partial _\alpha }{A_\beta }} \right){\delta _{\alpha \beta }}\text{d}{x_1} \wedge \text{d}{x_2} \cdots  \wedge \text{d}{x_d} \nonumber \\
 &= \left( {{\partial _\alpha }{A_\beta }} \right)\frac{1}{{\left( {d - 1} \right)!}}{\epsilon _{\beta {\nu _2} \cdots {\nu _d}}}\left( {{\epsilon _{\alpha {\nu _2} \cdots {\nu _d}}}\text{d}{x_1} \wedge \text{d}{x_2} \cdots  \wedge \text{d}{x_d}} \right) \nonumber \\
 &= \frac{1}{{\left( {d - 1} \right)!}}{\epsilon _{\beta {\nu _2} \cdots {\nu _d}}}\left( {{\partial _\alpha }{A_\beta }} \right)\text{d}{x_\alpha } \wedge \text{d}{x_{{\nu _2}}} \cdots  \wedge \text{d}{x_{{\nu _d}}} \nonumber \\
 &= \text{d} \left[ {\frac{1}{{\left( {d - 1} \right)!}}{\epsilon _{\beta {\nu _2} \cdots {\nu _d}}}{A_\beta }\, \text{d}{x_{{\nu _2}}} \cdots  \wedge \text{d}{x_{{\nu _d}}}} \right] .
\end{align}
Therefore, the Hodge dual picture makes it clear that counting scalar operators which are not total derivatives amounts to counting the number of non-exact \(d\)-forms:
\begin{equation}
 \# (\text{Indep scalar ops}) = \# (\text{non-exact } d \text{-form}) = \# (d\text{-form}) - \# ( \text{exact }d\text{-form}) .
\end{equation}
Since each exact $k$-form comes from a non-exact $(k-1)$-form, we arrive at the sequence truncated by the spacetime dimension:
\begin{align}
\# (\text{non-exact } d \text{-form}) & = \# (d \text{-form}) - \# (\text{non-exact } (d-1) \text{-form}) \nonumber \\
& = \# (d \text{-form}) - \left[\# ( (d-1) \text{-form}) - \# (\text{non-exact } (d-2) \text{-form}) \right] \nonumber \\
& = \cdots \nonumber \\
& = \sum\limits_{k=0}^d (-1)^k \# ( (d-k) \text{-form}) . \label{eq:sequence}
\end{align}
The number of each \(SO(d)\) rank $k$-form above can be easily projected out using character orthonormality $\int d\mu_\text{Lorentz}^{} \chi_a^* \chi_b^{} =\delta_{ab}$. In particular, to appropriately count the \(k\)-forms, under the Lorentz \(SO(d)\) integral we insert
\begin{align}
\sum_{k=0}^{d}(-1)^k\scD^k\ch_{(d-k)\text{-form}}
= \sum_{k=0}^{d}(-1)^k\scD^k\text{Tr}_{\Box}(\wedge^kg) = \text{det}_{\Box}(1-\scD g) 
\label{eq:sum_kforms}
\end{align}
where we have used the fact the \(k\)-form representation is obtained by the \(k\)-th exterior (antisymmetric) product of the vector (\(\Box\)) representation. 
We see that this corresponds to the \(1/P\) factor, eq.~\eqref{eq:1overP_def}, that enters into the Hilbert series integral in eq.~\eqref{eq:HSSM}. We understood eq.~\eqref{eq:1overP_def} to arise as a result of integrating over the conformal group; here we get a more operational understanding of how it is counting total derivative relations. Explicitly, and for future reference, we record here the result in four dimensions using \(SO(4) = SU(2)_L\times SU(2)_R\),
\begin{equation}
1 - \scD(\a + \a^{-1})(\b + \b^{-1}) + \scD^2(\a^2+\a^{-2} + \b^2+\b^{-2}+2) - \scD^3(\a + \a^{-1})(\b + \b^{-1}) + \scD^4 = \frac{1}{P(\scD,\a,\b)}.
\label{eq:1overP_4d}
\end{equation}

Although the sequence eq.~\eqref{eq:sequence} nicely gives the $1/P$ factor in eq.~\eqref{eq:HSSM}, we emphasize that it is not yet fully correct in accounting for the IBP relation. There are exceptions to the rule ``an exact $k$-form comes from a non-exact $(k-1)$-form'': sometimes even though a $(k-1)$-form is non-exact, its exterior derivative could still identically vanish, due to EOM or simply that it is a constant. In these exceptional cases, the $(k-1)$-form gives no $k$-form. In order to avoid over counting, one needs to carefully track these exceptional cases and make a corresponding fix, which amounts to including the  $\Delta H$ term in eq.~\eqref{eq:HSSM} (and given explicitly below for the SM).

\section{Hilbert series for the SM EFT}
\label{sec:HSSM}

In this section we discuss each of the elements of eq.~\eqref{eq:HSSM} explicitly for application to the SM EFT. We then explain how operators are counted in mass dimension, provide an example calculation for the SM Hilbert series, discuss how one can use the Hilbert series to aid in obtaining the explicit form of higher dimension operators, and show how to include multiple fermion flavors.

\subsection*{Elements of eq.~\eqref{eq:HSSM}}

We work in Euclidean space with Lorentz group \(SO(4) \simeq SU(2)_L \times SU(2)_R\). Representations are labeled by \((j_1,j_2)\); those needed are for scalars \((0,0)\), left-handed fermions \((\frac{1}{2},0)\) and their right-handed conjugates \((0,\frac{1}{2})\), and field strengths $X_{L/R}=\frac{1}{2}(X\pm \widetilde{X})$ in the $(1,0)$ and $(0,1)$ representations.

The field content of the SM is listed in Table~\ref{tbl:sm_fields}. Spurions in the Hilbert series are labeled by their field name, although we drop the subscript ``\(c\)'' for the left-handed conjugate fields \(u_c\), \(d_c\), and \(e_c\). That is, the Hilbert series is a function of the variables (for one fermion generation),
\be
H(\scD,\{\ph_a\})=H(\scD,Q,Q^\dag,L,L^\dag,H,H^\dag,u,u^\dag,d,d^\dag,e,e^\dag,B_L,B_R,W_L,W_R,G_L,G_R) .
\ee

In addition to the spacetime symmetry group, we impose invariance under the SM gauge group. Computations occur on the tori of these groups; we will use \(\a\) and \(\b\) to parameterize the torus of \(SU(2)_L\times SU(2)_R\), \(z_1\) and \(z_2\) for \(SU(3)_c\), \(y\) for \(SU(2)_W\), and \(x\) for \(U(1)_Y\).

\begin{table}
\centering
\[
\begin{array}{c|cc|ccc}
& SU(2)_L & SU(2)_R & SU(3)_c & SU(2)_W & U(1)_Y \\
\hline
H & \mathbf{1} & \mathbf{1} & \mathbf{1} & \mathbf{2} & 1/2 \\
Q & \mathbf{2} & \mathbf{1} & \mathbf{3} & \mathbf{2} & 1/6 \\
u_c & \mathbf{2} & \mathbf{1} & \xoverline{\mathbf{3}} & \mathbf{1} & -2/3 \\
d_c & \mathbf{2} & \mathbf{1} & \xoverline{\mathbf{3}} & \mathbf{1} & 1/3 \\
L & \mathbf{2} & \mathbf{1} & \mathbf{1} & \mathbf{2} & -1/2 \\
e_c & \mathbf{2} & \mathbf{1} & \mathbf{1} & \mathbf{1} & 1 \\
G_L & \mathbf{3} & \mathbf{1} & \mathbf{8} & \mathbf{1} & 0 \\
W_L & \mathbf{3} & \mathbf{1} & \mathbf{1} & \mathbf{3} & 0 \\
B_L & \mathbf{3} & \mathbf{1} & \mathbf{1} & \mathbf{1} & 0 \\
\end{array}
\]
\caption{\label{tbl:sm_fields} The SM field content and their charges under Lorentz and gauge groups. For each \(\ph_a\), the hermitian conjugate field \(\ph_a^{\dag}\) is also to be included.}
\end{table}

For a field \(\ph_a\), the character that enters the argument of the plethystic exponential is given by eq.~\eqref{eq:char_field},
\begin{equation}
\chi_a(\scD,\a,\b,x,y,z_1,z_2) = \chi^{}_{[d_a,(j_{1},\, j_{2})_a]}\,\chi^{\text{gauge}}_{a},
\end{equation}
where \(\chi^{}_{[\D,(j_1,j_2)]}\) is the character of the conformal group in four dimensions for a representation of scaling dimension \(\D\) and spin \((j_1,j_2)\). The characters necessary for the SM are
\begin{subequations}
\label{eq:char}
\begin{align}
\chi^{}_{[1,(0,0)]}(\scD,\alpha,\beta) =\ & \scD^{\textcolor{white}{1}} P(\scD, \a,\b)  (1-\scD^2) \\
\chi^{}_{[\frac{3}{2},(\frac{1}{2},\,0)]}(\scD,\alpha,\beta) =\ & \scD^{\frac{3}{2}}P(\scD, \a,\b)  \left(\a + \frac{1}{\a}-\scD \,\left(\b+\frac{1}{\b}\right) \right)\\
\chi^{}_{[\frac{3}{2},(0,\,\frac{1}{2})]}(\scD,\alpha,\beta) =\ & \scD^{\frac{3}{2}} P(\scD, \a,\b)  \left(\b + \frac{1}{\b}-\scD \,\left(\a+\frac{1}{\a}\right) \right)\\
\chi^{}_{[2,(1,0)]}(\scD,\alpha,\beta) =\ & \scD^2 P(\scD, \a,\b)  \left(\a^2+1+\frac{1}{\a} -\scD\left(\a+\frac{1}{\a}\right)\left(\b+\frac{1}{\b}\right)+\scD^2\right) \\
\chi^{}_{[2,(0,1)]}(\scD,\alpha,\beta) =\ & \scD^2 P(\scD, \a,\b)  \left(\b^2+1+\frac{1}{\b} -\scD\left(\b+\frac{1}{\b}\right)\left(\a+\frac{1}{\a}\right)+\scD^2\right)  \,,
\end{align}
\end{subequations}
The function \(P\) is
\begin{equation}
P(\scD,\a,\b) = \frac{1}{(1-\scD \a\b)(1-\scD/\a\b)(1-\scD\a/\b)(1-\scD\b/\a)}.
\end{equation}
The conformal characters in eq.~\eqref{eq:char} correspond to free fields with the EOM removed from the descendant operators~\cite{Barabanschikov:2005ri,Dolan:2005wy}. Note that the multiplicative \(\scD^{\D}\) factors in eq.~\eqref{eq:char} cancels against the factor \(\scD^{-d_a}\) that enters the argument of the plethystic exponential in eq.~\eqref{eq:HSSM}.

The character for the gauge group in the SM is $\chi^{\text{gauge}}_{R}=\chi^{U(1)}_R\chi^{SU(2)}_{R}\chi^{SU(3)}_{R}$, where the characters for the representations needed in the SM are
\begin{gather}
\chi^{U(1)}_Q(x) = x^Q \\
\chi^{SU(2)}_{{\bf 2}}(y) =\chi^{SU(2)}_{{\bf \overline{2}}}(y)= y+\frac{1}{y},~~~\chi^{SU(2)}_{\mathbf{ad}}(y)=  y^2+1+\frac{1}{y^2} \\
\chi^{SU(3)}_{{\bf 3}}(z_1,z_2) = z_1+\frac{z_2}{z_1}+\frac{1}{z_2}, ~~~~\chi^{SU(3)}_{{\bf \overline{3}}}(z_1,z_2) = z_2+\frac{z_1}{z_2}+\frac{1}{z_1}\\
\chi^{SU(3)}_{\mathbf{ad}}(z_1,z_2)=  z_1z_2+\frac{z_2^2}{z_1}+\frac{z_1^2}{z_2} + 2 + \frac{z_1}{z_2^2}+\frac{z_2}{z_1^2}+\frac{1}{z_1z_2}
\end{gather}

For example, the character of the spurion $Q$ in the SM is, 
\be
\chi^{}_Q =  \chi_{[\frac{3}{2},(\frac{1}{2},\,0)]}(\scD,\a,\b)\,\, \chi^{U(1)}_{1/6}(x) \,\,\chi^{SU(2)}_{\bf 2}(y) \,\, \chi^{SU(3)}_{\bf 3}(z_1,z_2) \,,\nonumber 
\ee
and the plethystic exponential is
\be
\text{PE}\Big[\frac{Q}{\scD^{\frac{3}{2}}}\chi^{}_Q\Big]= \exp\bigg(\sum_{r=1}^\infty && \, \frac{1}{r} (-1)^{r+1}\,\, \frac{Q^r}{\scD^{\frac{3}{2}r}} \,\, \chi^{}_{[\frac{3}{2},(\frac{1}{2},\,0)]}(\scD^r;\a^r,\b^r)\,\,\\
&&\times \chi^{U(1)}_{\frac{1}{6}}(x^r) \,\,\chi^{SU(2)}_{\bf 2}(y^r) \,\, \chi^{SU(3)}_{\bf 3}(z_1^r,z_2^r)  \bigg) \,.  \nonumber 
\ee

The measures of the integrals in eq.~\eqref{eq:HSSM} are the invariant group measures (Haar measures), normalized such that \(\int d\m^{}_G = 1\). For an integral \(\int d\m^{}_G \, f(g)\) with \(f(g)\) a class function, we can restrict the integration to the maximal torus using the Weyl integration formula.\footnote{We are unaware of a great physics-oriented reference for this. Weyl's work~\cite{Weyl:classical} might actually be the most straightforward to understand. We will attempt to give a readable derivation in~\cite{Henning:2017fpj}. Most math textbooks on group theory cover the Weyl integration formula; one that we like is~\cite{Brocker:2003}. Formulas for the present Haar measures can be found in the, \textit{e.g.}, the physics paper~\cite{Gray:2008yu}} The Haar measures we give in the following are written in this way. For the Lorentz group
\small
\be
\int d\mu_{SU(2)_L\times SU(2)_R} = \frac{1}{4}\frac{1}{(2\pi i)^2}\oint_{|\a|=1} \oint_{|\b|=1}\frac{d \a}{\a} \frac{d\b}{\b}  \,\,\left(1-\alpha ^2\right) \big(1-\frac{1}{\alpha^{2}}\big)\left(1-\beta ^2\right)\big(1-\frac{1}{\beta^{2}}\big),
\ee
\normalsize
where $\a$ is the parameter associated with $SU(2)_L$ and $\b$ with $SU(2)_R$.
For the SM gauge group we have
\be
\int d\mu_{\text{gauge}} = \int d\mu_{U(1)} \,\int d\mu_{SU(2)}\,\int d\mu_{SU(3)} \,,
\ee
with
\small
\begin{subequations}
\be
\int d\mu_{U(1)} &=& \frac{1}{2\pi i}\oint_{|x|=1} \frac{dx}{x} , \\
\int d\mu_{SU(2)} &=& \frac{1}{2}\frac{1}{2\pi i}\oint_{|y|=1} \frac{dy}{y} \,\,\left(1-y^2\right) \big(1-\frac{1}{y^{2}}\big) , \\
\int d\mu_{SU(3)} &=& \frac{1}{6}\frac{1}{(2\pi i)^2}\oint_{|z_1|=1} \oint_{|z_2|=1} \frac{dz_1}{z_1} \frac{dz_2}{z_2} \big(1-z_1z_2\big)\big(1-\frac{z_1^2}{z_2}\big) \big(1-\frac{z_2^2}{z_1}\big) \big(1-\frac{1}{z_1z_2}\big) \nonumber \\ &&\qquad \qquad \qquad \qquad \qquad \qquad \qquad \times \big(1-\frac{z_2}{z_1^2}\big) \big(1-\frac{z_1}{z_2^2}\big) \, .
\ee
\end{subequations}
\normalsize

Finally, the $\Delta H$ modification term in eq.~\eqref{eq:HSSM} can be obtained from the general formula presented in \cite{Henning:2017fpj}. Explicitly, for the SM (with one fermion generation) it is given by
\be
\Delta H(\scD,\{\ph_a\}) =
d d^\dag \scD + e e^\dag \scD + L L^\dag \scD + Q Q^\dag \scD +
  u u^\dag \scD - B_L \scD^2 - B_R \scD^2 + H H^\dag \scD^2 - \scD^4  \,. \nonumber \\
\label{eq:delH}
\ee
Note that all of these terms are non-sensical when interpreted as operators; their job is to cancel such terms coming from the remainder of eq.~\eqref{eq:HSSM}.

\subsection*{Counting in mass dimension}
To obtain the counting of SM operators at a given mass dimension, we re-scale the spurions according to their mass dimension: 
\be
\begin{split}
 \scD \to&\,\, \e\scD   \,, \\
 \ph_a \to&\,\, \e^{d_a}\ph_a \,,
 \end{split}
 \label{eq:scaling}
 \ee
 with \(d_a\) the canonical scaling dimension of the field \(\ph_a\). The Hilbert series eq.~\eqref{eq:hil_def} can then be viewed as a series in mass dimension,
\begin{equation}
H(\e,\scD,\{\ph_a\}) = \sum_{r_1,\dots,r_N}\sum_{k} c_{\mathbf{r}\,k}\, \e^{\mathbf{d}\cdot\mathbf{r}+ k}\ph_1^{r_1}\dots\ph_N^{r_N}\scD^k \equiv \sum_{i}\e^i \widehat{H}_i(\scD,\{\ph_a\}).
\label{eq:hil_def_mass}
\end{equation}
Recalling eq.~\eqref{eq:HSSM}, let us define
\be
H^0( \scD,\{ \phi_a\}) = \int d\mu_{\text{Lorentz}}\int  d \mu_{\text{gauge}} \,\,\frac{1}{P(\scD,\a,\b)} \,\,\text{PE}\Big[\sum_a \frac{\ph_a}{\scD^{d_a}}\ch_a\Big]  \,,
\label{eq:master}
\ee
such that eq.~\eqref{eq:HSSM} takes the form $H = H^0+\Delta H$. 
The terms in the Hilbert series of mass dimension four and less are given by
\be
\begin{split}
\sum_{i=0}^4 \e^i \, \widehat{H}_i  =&\,\, H^0( \e \, \scD, \{\e^{d_a} \phi_a \})\bigg|_{\epsilon\le4}+  \Delta H( \e \, \scD, \{\e^{d_a} \phi_a \})  \\ 
=&\,\, 1+\e^2 H H^\dagger  + \e^4\bigg( B_L^2 + B_R^2 + W_L^2 + W_R^2 + G_L^2 + G_R^2 + H^2 {H^\dagger}^2   \\
~& +\,e H^{\dagger} L + e^\dagger H L^\dag + d H^\dag Q + d^\dag H Q^\dag + H Q u + H^\dag Q^\dag u^\dag \bigg) \,,
\end{split}
\label{eq:hle4}
\ee
where the notation $|_{\epsilon\le4}$ means taking a Taylor expansion of the integrand of eq.~\eqref{eq:master} in powers of $\e$ up to order $\e^4$, and where $\Delta H$ was given in eq.~\eqref{eq:delH}. At mass dimension five and above, we have
\be
i\ge 5: ~~~~ \widehat{H}_i =H^0(\e \scD,\{\e^{d_a}\phi_a\})\bigg|_{\mathcal{O}(\e^i)} \,, 
\label{eq:hge5}
\ee
where the notation $|_\mathcal{O}(\epsilon^i)$ means taking the  coefficient of $\e^i$ in the Taylor expansion of the integrand of eq.~\eqref{eq:master} in powers of $\e$.
In the Taylor expansion, all poles are located at the origin. Because all poles are at 0, the computation is very straightforward and simple to program.
We see explicitly in eqs.~\eqref{eq:hle4},~\eqref{eq:hge5} that $\Delta H$ does not affect mass dimension five and above.

Note that:  {\it i)}  the Hilbert series does {\it not} include kinetic terms for scalar and fermion fields, since in our  method these are counted as zero on account of the EOM (these terms can of course be trivially added in to eq.~\eqref{eq:hle4} such that it accounts for all the terms in the usual `renormalizable' SM Lagrangian); {\it ii)} because we work with field strength tensors (and not gauge fields), the gauge kinetic terms {\it are} already present in the Hilbert series; and, {\it iii)} the fields $\tilde F$, $\tilde W$ and $\tilde G$ are used as building blocks for the Lagrangian, so that topological terms such as $F\tilde F$ {\it are}  included in the Hilbert series.

\subsection*{Example outputs: dimension 5 and 6 Hilbert series}
We can expand the integrand in eq.~\eqref{eq:master} in $\epsilon$ and retain only the coefficient of $\epsilon^5$; performing the contour integrals over the Lorentz and gauge group parameters, we pick up residues at 0 in all the variables  $\a,\b,x,y,z_1,z_2$,
\small
\be
&&H_{\text{SM}}\bigg|_{O(\epsilon^5)}\equiv \widehat{H}_5 =\nonumber \\
 &=&\oint \frac{d\a}{2\pi i}  \oint \frac{d\b}{2\pi i} \frac{\left(1-\alpha ^2\right)^2 \left(1-\beta ^2\right)^2}{4 \alpha ^3 \beta
   ^3} 
 \oint \frac{dx}{2\pi i}\frac{1}{x}
 \oint \frac{dy}{2\pi i}\frac{\left(1-y^2\right)^2}{2 y^3}
 \oint \frac{dz_1}{2\pi i} \oint \frac{dz_2}{2\pi i}
 \frac{\left(z_1^2-z_2\right)^2 (1-z_1z_2)^2
   \left(z_1-z_2^2\right)^2}{6 z_1^5 z_2^5}
     \nonumber \\ &&\bigg[  H^2 L^2 \frac{\left(y^4+y^2+1\right) \left(\alpha ^2+\alpha ^2 y^4+\left(\alpha
   ^2+1\right)^2 y^2\right)}{\alpha ^2 y^4}
 +  (H^\dag)^2 (L^\dag)^2 \frac{\left(y^4+y^2+1\right) \left(\beta ^2+\beta ^2 y^4+\left(\beta
   ^2+1\right)^2 y^2\right)}{\beta ^2 y^4} + \ldots \bigg]  \nonumber \\
   &=&H^2 L^2 +H^{\dag\,2} L^{\dag\,2} \,,
\ee
\normalsize
where the $+\ldots$ inside the large brackets are terms which evaluate to zero upon performing the contour integrals. This is the Hilbert series for dimension-five operators in the SM EFT. One readily identifies that the Hilbert series is picking up the well known operators which give neutrino masses.

Repeating this at order $\epsilon^6$ we obtain the Hilbert series for dimension-six operators of the SM EFT:
\small
\be
\widehat{H}_{6}&&=H^{3}H^{\dag\,3}+u^{\dag}Q^{\dag}HH^{\dag\,2}+2Q^{2}Q^{\dag\,2}+Q^{\dag\,3}L^{\dag}+Q^{3}L+2QQ^{\dag}LL^{\dag}+L^{2}L^{\dag\,2}+uQH^{2}H^{\dag} \nonumber \\ && \,\,
+\,2uu^{\dag}QQ^{\dag}+uu^{\dag}LL^{\dag}+u^{2}u^{\dag\,2}+e^{\dag}u^{\dag}Q^{2}+e^{\dag}L^{\dag}H^{2}H^{\dag}+2e^{\dag}u^{\dag}Q^{\dag}L^{\dag}+eLHH^{\dag\,2}+euQ^{\dag\,2} \nonumber \\ &&\,\,
+\,2euQL+ee^{\dag}QQ^{\dag}+ee^{\dag}LL^{\dag}+ee^{\dag}uu^{\dag}+e^{2}e^{\dag\,2}+d^{\dag}Q^{\dag}H^{2}H^{\dag}+2d^{\dag}u^{\dag}Q^{\dag\,2}+d^{\dag}u^{\dag}QL \nonumber \\ &&\,\,
+\,d^{\dag}e^{\dag}u^{\dag\,2}+d^{\dag}eQ^{\dag}L+dQHH^{\dag\,2}+2duQ^{2}+duQ^{\dag}L^{\dag}+de^{\dag}QL^{\dag}+deu^{2}+2dd^{\dag}QQ^{\dag}+dd^{\dag}LL^{\dag} \nonumber \\ &&\,\,
+\,2dd^{\dag}uu^{\dag}+dd^{\dag}ee^{\dag}+d^2d^{\dag\,2}+u^{\dag}Q^{\dag}H^{\dag}G_R+d^{\dag}Q^{\dag}HG_R+HH^{\dag}G_R^{2}+G_R^{3}+uQHG_L \nonumber \\ &&\,\,
+\,dQH^{\dag}G_L+HH^{\dag}G_L^{2}+G_L^{3}+u^{\dag}Q^{\dag}H^{\dag}W_R+e^{\dag}L^{\dag}HW_R+d^{\dag}Q^{\dag}HW_R+HH^{\dag}W_R^{2}+W_R^{3} \nonumber \\ &&\,\,
+\,uQHW_L+eLH^{\dag}W_L+dQH^{\dag}W_L+HH^{\dag}W_L^{2}+W_L^{3}+u^{\dag}Q^{\dag}H^{\dag}B_R+e^{\dag}L^{\dag}HB_R \nonumber \\ &&\,\,
+\,d^{\dag}Q^{\dag}HB_R+HH^{\dag}B_RW_R+HH^{\dag}B_R^{2}+uQHB_L+eLH^{\dag}B_L+dQH^{\dag}B_L+HH^{\dag}B_LW_L \nonumber \\ &&\,\,
+\,HH^{\dag}B_L^{2}+2QQ^{\dag}HH^{\dag}\mathcal{D}+2LL^{\dag}HH^{\dag}\mathcal{D}+uu^{\dag}HH^{\dag}\mathcal{D}+ee^{\dag}HH^{\dag}\mathcal{D}+d^{\dag}uH^{2}\mathcal{D}+du^{\dag}H^{\dag\,2}\mathcal{D} \nonumber \\ &&\,\,
+\,dd^{\dag}HH^{\dag}\mathcal{D}+2H^{2}H^{\dag\,2}\mathcal{D}^{2} \,.
\label{eq:dim6}
\ee
\normalsize
Setting all of the spurions equal to unity gives $\widehat{H}_6=84$, the total number of independent local operators at dimension 6, but more information is contained in eq.\eqref{eq:dim6}. For instance, the counting can easily be further decomposed by baryon number violation, $76+8$. The perhaps more familiar `$59+4$' counting is one in which hermitian conjugates of fermionic operators are not counted separately (such counting can of course also be obtained from eq.~\eqref{eq:dim6}). 

\subsection*{Explicit form of the operators}
At low dimensions (including dimension 7 and 8), explicitly constructing an operator basis requires minimal effort. For example, the $+Q^3 L$ term in eq.~\eqref{eq:dim6} tells us that there is one independent operator composed of three powers of $Q$ and one power of $L$; the $+2L L^\dag Q Q^\dag$ term that there are two independent operators composed of one power each of $L,L^\dag,Q$ and $Q^\dag$; the $+2H H^\dag Q Q^\dag \mathcal{D}$ term that there are two independent operators composed of $H, H^\dag, Q, Q^\dag$ and one covariant derivative, \textit{etc}. Exactly how derivatives act and how Lorentz and gauge indices are contracted is information beyond what the Hilbert series can provide. However, such information can be easily deduced for low-order terms. For example, in the $2H H^\dag Q Q^\dag \mathcal{D}$ term, because the combination $Q Q^\dag$ has to be formed into a Lorentz singlet, it follows there must be a $\slashed{D}=\gamma^\mu D_\mu$, \textit{i.e.} ${\bar Q}\gamma^\mu Q D_\mu$; the gauge indices can be contracted in two inequivalent ways: $i \left[H^\dagger (D_\mu H) - (D_\mu H^\dagger) H\right] {\bar Q}\gamma^\mu Q$ and $i\left[H^\dagger \t^a (D_\mu H) - (D_\mu H^\dagger) \t^a H\right] {\bar Q}\gamma^\mu \t^a Q$, where \(\t^a\) are the \(SU(2)_W\) generators.

\subsection*{Multiple flavors}
The inclusion of additional fermion families is trivial---simply add the extra fields into the PE. Alternatively the PE of each fermion family can be raised to the power of $N_f$---the results we selected to show below use this counting for ease of display, but in doing this additional information about the flavor structure is missing compared to the case where one distinguishes between the fermion families.

\section{Selected results}
\label{sec:results}

Full Hilbert series for the SM EFT up to mass dimension 12 are supplied as an auxiliary {\tt Mathematica} file to this paper. In this section we present selected results. Our convention is to separately count operators which are related by hermitian conjugation; this counts all CP-even and CP-odd operators independently.

\subsection{Operator bases at dimension 7 and 8}

In this section we revisit the SM EFT operator basis at dimension 7 and 8 for arbitrary number of flavors \(N_f\). We find that the existing analysis~\cite{Lehman:2015coa}---which includes full $N_f$ dependence at dimension 7, and $N_f=1$ at dimension 8---missed some operators containing two or more derivatives. In the present section we will explain why that analysis missed some operators and then summarize the dim-7 and dim-8 operator content. Appendix~\ref {app:dim8_comparison} contains a comprehensive listing and analysis of the operators missing from~\cite{Lehman:2015coa}.

From a strictly computational point of view, the only difference in our method compared to~\cite{Lehman:2015coa} is the \(1/P\) factor in eqs.~\eqref{eq:HSSM} and~\eqref{eq:master}. Operationally, the \(1/P\) factor accounts for IBP relations; in section~\ref{subsec:HodgeDual}, we presented an understanding of this using the language of differential forms: in four dimensions, the independent IBP relations come from non-exact 3-forms. Hence, the correct counting takes the number of 4-forms (scalars) and subtracts the number of non-exact 3-forms. In~\cite{Lehman:2015coa}, the authors take the number of 4-forms and subtract \textit{all} (exact and non-exact) 3-forms (\textit{i.e.}, the first two terms in eq.~\eqref{eq:1overP_4d}); they point out that this gives spurious results but were unable to find a systematic procedure to correct it. Here we find that correct fix is to add in the rest of the terms of eq.~\eqref{eq:1overP_4d}; this makes it clear that the differences we find occur when there are two or more derivatives. 

Moving to results, we follow~\cite{Lehman:2015coa} and group the counting of operators into classes which do not mix under EOM or IBP. In the SM EFT we count classes by {\it i)} number of derivatives $\mathcal{D}$, {\it ii)} powers of Higgs field $H$ (or $H^\dag$), {\it iii)} powers of gauge field $X$ (i.e. $X\in \{B_L,B_R,W_L,W_R,G_L,G_R\}$),  {\it iv)} powers of fermion fields $\psi$ (i.e. $\psi \in \{Q,u,d,L,e\}$ and their conjugates).

\subsubsection*{Dimension 7}
\begin{table}\footnotesize
\renewcommand\arraystretch{1.4}
\[
\begin{array}{cc|ccrr}
\text{Class} & & N_f & & 1 & 3\\
\hline
\hline
XH^2\ps^2 & & N_f(3N_f - 1) & & 2 & 24 \\
H^4 \psi^2 & & N_f(N_f+1) & & 2 & 12\\ 
\multirow{2}{*}{$H\ps^4 $} & (B) &  \frac{2}{3} N_f^2(14N_f^2 + 1)  & &  10  &  762 \\
 & (\slashed{B}) &  \frac{1}{3}N_f^2(17N_f^2 -3N_f -2)  & &  4 &  426  \\
\hline
H^3\ps^2\scD &  & 2N_f^2 & & 2 & 18 \\
\multirow{2}{*}{$\ps^4\scD$} & (B) & N_f^3(N_f+1)& & 2 &  108 \\
 & (\slashed{B}) & \frac{1}{3}N_f^2(4N_f^2 + 6N_f + 2) & & 4 & 168 \\
\hline
H^2\ps^2\scD^2 & & 2N_f(N_f+1) & & 4 & 24 \\
\hline
\hline
\multirow{2}{*}{Total} & (B) & \frac{31}{3}N_f^4 + N_f^3 + \frac{26}{3} N_f^2 + 2N_f & & 22 & 948 \\
& (\slashed{B}) & 7N_f^4 + N_f^3 & & 8 & 594
\end{array}
\]
\caption{\label{tbl:dim7} Dimension 7 operators for arbitrary \(N_f\) as well as for \(N_f=1\) and \(N_f=3\). Lines separate classes involving no derivatives, one derivative, and two derivatives. Operators involving four fermions are further distinguished either as preserving baryon number \((B)\) or violating baryon number \((\slashed{B})\). }
\end{table}

The different classes of dimension-seven operators and the number of operators in them are summarized in Table~\ref{tbl:dim7}. In total, we find 30 operators for \(N_f=1\) and 1542 for \(N_f=3\). All dim-7 operators violate either lepton or baryon number; moreover, all violate \(B-L\). The dim-7 operators have either zero, one, or two derivatives. As the additional operators we find compared to~\cite{Lehman:2015coa} all involve two or more derivatives, there is only one term at conflict between our analyses. For the two derivative class \(H^2\ps^2\scD^2\) we find:
\begin{equation}
\widehat{H}_7 \supset N_f(N_f+1)H^2L^2\scD^2 + \text{h.c.} \ .
\end{equation}
This is a larger coefficient than that found in~\cite{Lehman:2015coa} (that analysis produced a coefficient of \(N_f(N_f+3)/2\), which numerically differs from ours for \(N_f>1\)). 

\subsubsection*{Dimension 8} 
\begin{table}\footnotesize
\renewcommand\arraystretch{1.4}
\[
\begin{array}{cc|ccrr}
\text{Class} & & N_f & & 1 & 3\\
\hline
X^4 & & 43 & & 43 & 43\\
X^3 H^2 & & 6 & & 6  & 6\\
X^2 H^4 & & 10 & & 10 & 10\\
H^8 & & 1 & & 1 & 1\\
X^2H \ps^2 & & 96N_f^2 & & 96 & 864\\
XH^3 \ps^2 & & 22N_f^2 & & 22 & 198\\
H^5 \ps^2 & & 6N_f^2 & & 6  & 54\\
\multirow{2}{*}{$X \ps^4$} & (B)& 4N_f^2(40N_f^2 - 1) & & 156 & 12924\\
& (\slashed{B}) & 2N_f^3(21N_f +1) & & 44 & 3456\\
\multirow{2}{*}{$H^2\ps^4$} & (B) & N_f^2(67N_f^2 +N_f + 7) & & 75 & 5517 \\
& (\slashed{B}) & \frac{1}{3}N_f^2(43N_f^2 - 9N_f + 2) & & 12 & 1086 \\
\hline
X^2\ps^2\scD & & 57N_f^2 & & 57 & 513\\
XH^2\ps^2\scD & & 92N_f^2 & & 92 & 828\\
H^4\ps^2\scD & & 13N_f^2 & & 13 & 117\\
\multirow{2}{*}{$H\ps^4\scD$} & (B) & N_f^3(135N_f-1) & & 134 & 10908\\
& (\slashed{B}) & N_f^3(29N_f+3) & & 32 & 2430\\
\hline
X^2H^2\scD^2 & & 18 & & 18 & 18\\
XH^4\scD^2 & & 6 & & 6 & 6\\
H^6\scD^2 & & 2 & & 2 & 2\\
XH\ps^2\scD^2 & & 48N_f^2 & & 48 & 432\\
H^3\ps^2\scD^2 & & 36N_f^2 & & 36 & 324\\
\multirow{2}{*}{$\ps^4\scD^2$} & (B) & \frac{11}{2}N_f^2(9N_f^2 + 1) & & 55 & 4059\\
& (\slashed{B}) & N_f^3(11N_f - 1) & & 10 & 864 \\
\hline
H^2\ps^2\scD^3 & & 16N_f^2 & & 16 & 144\\
\hline
H^4\scD^4 & & 3 & & 3 & 3\\
\hline
\hline
\multirow{2}{*}{Total} & (B) & \frac{823}{2}N_f^4 + \frac{789}{2} N_f^2 + 89 & & 895 & 36971 \\
& (\slashed{B}) & \frac{289}{3}N_f^4 + N_f^3 + \frac{2}{3} N_f^2  & & 98 & 7836
\end{array}
\]
\caption{\label{tbl:dim8} Dimension 8 operators with conventions as in Table~\ref{tbl:dim7}.}
\end{table}

Our analysis at dimension 8 is the first to include full $N_f$ dependence. The results are summarized in Table~\ref{tbl:dim8}. In total, we find 993 operators for \(N_f = 1\) and 44807 for \(N_f = 3\). We highlight:
\begin{itemize}
\item There are 62 additional operators in the $N_f=1$ case which were not uncovered in~\cite{Lehman:2015coa}. These operators involve two or three derivatives, and an explicit listing of them is given in Appendix~\ref {app:dim8_comparison}.
\item All baryon violating operators are \(\D B = 1\) and all preserve \(B-L\).
\item There are 10 types of operators which only appear for $N_f>1$, five of which are baryon number violating. Seven of these operators occur in the class \(X\ps^4\), while three occur in the class \(H^2\ps^4\). The baryon preserving ones are 
\small
\be
\widehat{H}_8 \supset && 
N_f^2 \left(N_f^2-1\right)Q^{2}Q^{\dag\,2}B_L,~\nonumber
\frac{1}{2} N_f^2 \left(N_f^2-1\right)L^{2}L^{\dag\,2}B_L,~\frac{1}{2} N_f^2 \left(N_f^2-1\right)u^{2}u^{\dag\,2}B_L,~\nonumber \\&&\frac{1}{4} N_f^2 \left(N_f^2-1\right)e^{2}e^{\dag\,2}B_L,~\frac{1}{2} N_f^2 \left(N_f^2-1\right)d^2d^{\dag\,2}B_L \ \ + \ \ \text{h.c.} \, ,
\ee 
\normalsize
while the baryon violating ones are
\small
 \be
\widehat{H}_8 \supset && \frac{1}{2} N_f^3(N_f-1)euQ^{\dag\,2}W_L,~\frac{1}{2} N_f^3(N_f-1) euQ^{\dag\,2}B_R,~\frac{1}{2} N_f^3(N_f-1) d^{\dag}e^{\dag}Q^{2}H^{2},~\nonumber \\ && \frac{1}{2} N_f^3(N_f-1) d^{\dag\,2}QLH^{2},~\frac{1}{2} N_f^3(N_f-1) u^{2}Q^{\dag}L^{\dag}H^{2} \ \ + \ \ \text{h.c.} \, .
\ee 
\normalsize
\end{itemize}

\subsection{Counting at higher dimensions with arbitrary $N_{f}$}
\begin{figure}
\centering
\includegraphics[width=13cm]{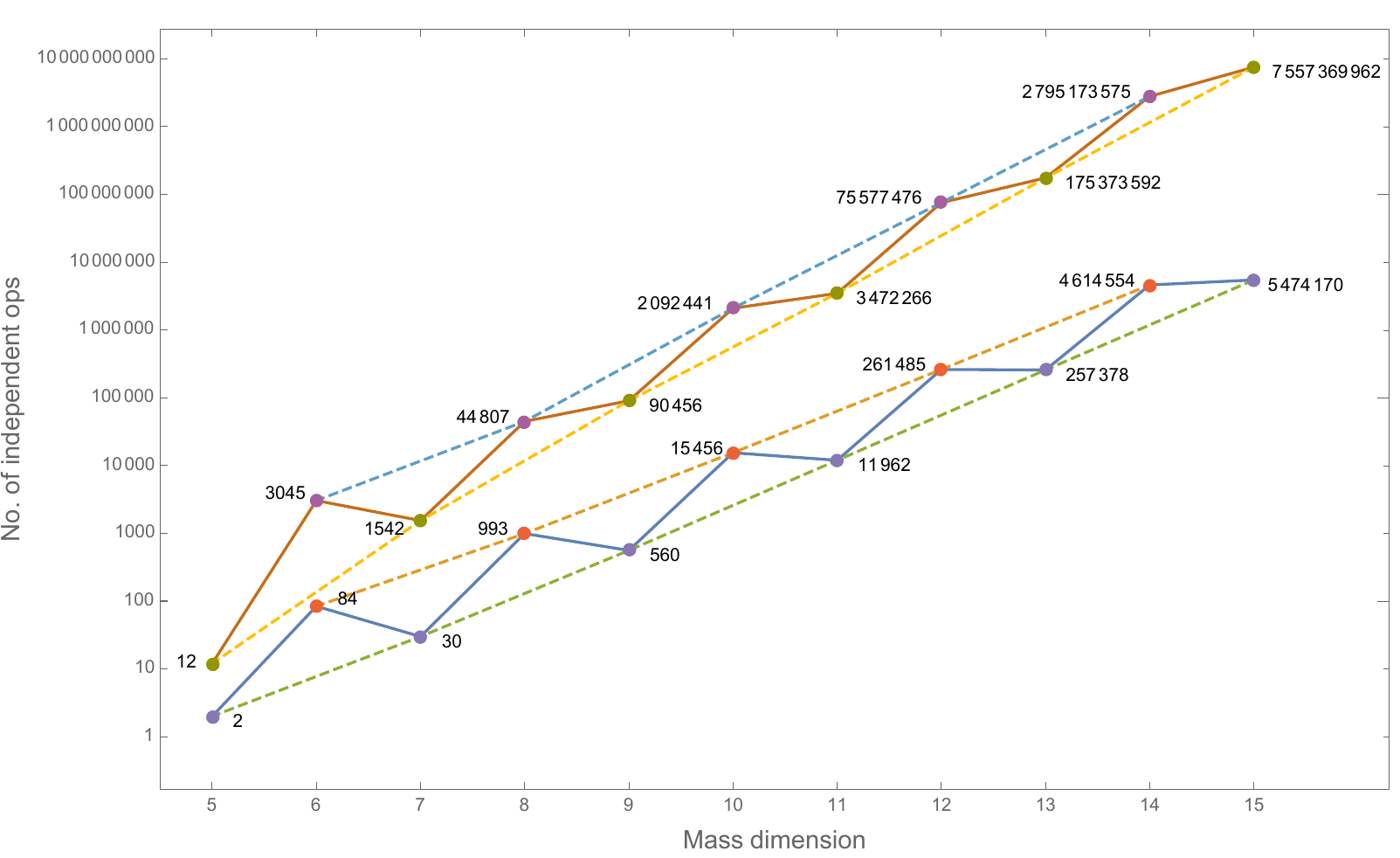}
\caption{Growth of the number of independent operators in the SM EFT up to mass dimension 15. Points joined by the lower solid line are for one fermion generation; those joined by the upper solid line are for three generations. Dashed lines are to guide the eye to the growth of the even and odd mass dimension operators in both cases.}
\label{fig:growth}
\end{figure}

Here we present the number of independent operators in the SM EFT as a function of $N_f$, up to mass dimension 15. The
counting is split up into $(\Delta B=0)+(\Delta B=1)+(\Delta B=2)$ parts ($B$ is baryon number) up to dimension 12, with the understanding that when there is only one bracket, it is $\Delta B=0$, and when there are only two, the first is $\Delta B=0$ and the second is $\Delta B=1$. We find,\footnote{A typographical error in copying the output of the Hilbert series into the $\Delta B=0$ term at mass dimension 12 has been corrected from the first version of this paper. (Previously the first set of parentheses at mass dimension 12 included the sum of both the $\Delta B=0$ and $\Delta B = 2$ terms.) We thank Renato Fonseca for bringing this to our attention.}
\small
\be
\text{\# Dim 5}&=&\bigg(N_f + N_f^2\bigg) \nonumber \\
\text{\# Dim 6}&=&\bigg(15 + \frac{135}{4}N_f^2 + \frac{1}{2}N_f^3 + \frac{107}{4} N_f^4\bigg) 
+\bigg(\frac{2}{3} N_f^2 + N_f^3 + \frac{19}{3} N_f^4\bigg) \nonumber \\
\text{\# Dim 7}&=&\bigg(2 N_f + \frac{26}{3} N_f^2 + N_f^3 + \frac{31}{3} N_f^4\bigg) +\bigg(N_f^3 + 7 N_f^4\bigg)  \nonumber \\
\text{\# Dim 8}&=&\bigg(89 + \frac{789}{2} N_f^2 + \frac{823}{2} N_f^4\bigg)+\bigg(\frac{2}{3} N_f^2 + N_f^3 + \frac{289}{3} N_f^4\bigg)\nonumber \\
\text{\# Dim 9}&=&\bigg(9 N_f + 83 N_f^2 + \frac{49}{12} N_f^3 + \frac{2587}{12} N_f^4 - \frac{1}{12}N_f^5 + \frac{437}{12} N_f^6\bigg)  \nonumber \\
&&+ ~\bigg(-\frac{4}{3} N_f^2 + \frac{29}{3} N_f^3 + \frac{463}{3} N_f^4 + \frac{1}{3}N_f^5 + 41 N_f^6 \bigg)
+ \bigg(\frac{1}{4}N_f^2 + \frac{61}{24} N_f^3 + \frac{29}{24} N_f^4 + \frac{11}{24} N_f^5 + \frac{85}{24} N_f^6 \bigg)\nonumber \\
\text{\# Dim 10}&=&  \bigg( 530 + \frac{53927}{12} N_f^2 - \frac{17}{2} N_f^3 + \frac{82127}{12} N_f^4 - 6 N_f^5 + \frac{3776}{3} N_f^6 \bigg) \nonumber \\
&&+~ \bigg(-\frac{10}{9} N_f^2 + \frac{155}{3} N_f^3 + \frac{30169}{18} N_f^4 + \frac{37}{3} N_f^5 + \frac{10891}{18} N_f^6\bigg) \nonumber \\
\text{\# Dim 11}&=&\bigg( 18 N_f + \frac{2812}{3} N_f^2 - \frac{152}{3} N_f^3 + \frac{11689}{3} N_f^4 - \frac{58}{3} N_f^5 + \frac{5551}{3} N_f^6  \bigg) \nonumber \\
&&+ ~\bigg(-2 N_f^2 + \frac{443}{3} N_f^3 + \frac{8830}{3} N_f^4 + \frac{352}{3} N_f^5 + \frac{5855}{3} N_f^6\bigg)  \nonumber \\
&&+ ~\bigg( \frac{3}{4} N_f^2 + \frac{307}{24} N_f^3 + \frac{7}{24} N_f^4 + \frac{197}{24} N_f^5 + \frac{3599}{24} N_f^6 \bigg)\nonumber \\
%
\text{\# Dim 12}&=& \bigg( 4481 + \frac{1}{2}N_f + \frac{613247}{12} N_f^2 - \frac{674}{3} N_f^3 + \frac{1743611}{16} N_f^4 - \frac{
 8965}{24} N_f^5 + \frac{132565}{3} N_f^6 - \frac{187}{24} N_f^7 + \frac{22137}{16} N_f^8 \bigg) \nonumber \\
&&+ ~\bigg( \frac{28}{9} N_f^2 + \frac{1954}{3} N_f^3 + 27779 N_f^4 + \frac{6823}{12} N_f^5 + \frac{131429}{6} N_f^6 + \frac{169}{12} N_f^7 + \frac{17803}{18} N_f^8  \bigg) \nonumber \\
&&+ ~\bigg( \frac{11}{24} N_f^3 + \frac{1483}{144} N_f^4 + \frac{19}{12} N_f^5+ \frac{149}{72} N_f^6 + \frac{47}{24} N_f^7 + \frac{4555}{144} N_f^8\bigg) \nonumber
\ee
\normalsize
The number 2499 of baryon conserving operators at dimension 6 with three generations \cite{Alonso:2013hga} is recovered. Although we only presented the numbers above, the full content of the Hilbert series is contained in the accompanying \texttt{Mathematica} file.
As the Hilbert series begin to become extremely lengthy, we continue counting operators without retaining the content information ({\it i.e.} setting all spurions equal to unity), but still retaining $N_f$ dependence:
\small 
\be
\text{\# Dim 13}&=&-109 N_f+\frac{159296}{15}N_f^2+\frac{32063 }{90}N_f^3+\frac{5140756}{45}N_f^4+\frac{78253}{72}N_f^5+\frac{42846881 }{360}N_f^6+\frac{68723}{360}N_f^7\nonumber \\
&&~~~+\frac{4311047}{360}N_f^8 \nonumber \\
\text{\# Dim 14}&=&40715-2 N_f+\frac{105860297}{180}N_f^2+\frac{89759 }{18}N_f^3+\frac{1513774187}{720}N_f^4+\frac{63971}{72}N_f^5+\frac{299553293}{180}N_f^6\nonumber \\
&&~~~-\frac{117979 }{72}N_f^7+\frac{51562231 }{240}N_f^8 \nonumber \\
\text{\# Dim 15}&=&-2427 N_f+\frac{21647887 }{180}N_f^2-\frac{114619}{20}N_f^3+\frac{387130705 }{216}N_f^4-\frac{10026269}{1440}N_f^5+\frac{456200951}{160}N_f^6\nonumber \\
&&~~~-\frac{3717991}{720}N_f^7+\frac{103741331}{144}N_f^8-\frac{534941}{1440}N_f^9 +\frac{9163865 }{864}N_f^{10} \nonumber 
\ee
\normalsize
(which exhibit some rather large prime numbers!). The number of independent operators evaluated for $N_f=1$ and $N_f=3$ up to dimension 15 are plotted in Fig.~\ref{fig:growth}. We see the growth is exponential, which is to be expected on general grounds~\cite{Cardy:1991kr}.

\section{Discussion}
\label{sec:discuss}

The method we have outlined in this paper can be extended trivially to determining the content and number of higher dimension operators for any four-dimensional relativistic gauge theory with scalar and fermionic matter. The master equation is eq.~\eqref{eq:master}, which needs to be modified from the SM to the theory of interest. The pieces of eq.~\eqref{eq:master} which are SM specific are the gauge groups (and as such the Haar measures that need to be integrated over to produce gauge singlets), and the field content (which enters the plethystic exponential).  

In the present work we studied the expansion of eq.~\eqref{eq:HSSM} in powers of mass dimension, $\epsilon$. However, in our previous work in (0+1) dimensions \cite{Henning:2015daa} we were able to obtain all-order formulae for Hilbert series, revealing a fascinating analytic structure which could not be seen in any finite order expansion. Can we hope to attack eq.~\eqref{eq:HSSM} directly? Could this reveal some previously hidden all-order structure of the SM EFT? While lofty, questions along these lines merit detailed investigation of the structure underlying operator bases, which we take up in~\cite{Henning:2017fpj}.

\acknowledgments
BH is grateful to Witold Skiba for conversations. BH supported in part by the U.S. DOE under the contract DE-FG02-92ER-40704. XL is supported by DOE grant DE-SC-000999.
TM is supported by U.S. DOE grant DE-AC02-05CH11231 and acknowledges  
computational resources provided through ERC grant number 291377: ``LHCtheory''. 
HM was supported by the U.S. DOE under Contract DE-AC02-05CH11231, and
by the NSF under grants PHY-1316783 and PHY-1638509.  HM was also
supported by the JSPS Grant-in-Aid for Scientific Research (C)
(No.~26400241 and 17K05409), MEXT Grant-in-Aid for Scientific Research on Innovative Areas (No. 15H05887, 15K21733), and by WPI, MEXT, Japan.


\appendix
\section{Explicit comparison with~\cite{Lehman:2015coa} at dim-8}
\label{app:dim8_comparison}

In this appendix we present the comparison with the results and method of Lehman and Martin (LM)~\cite{Lehman:2015coa}. As explained in section~\ref{sec:results}, differences will occur as, operationally, LM only use the first two terms in eq.~\eqref{eq:1overP_4d} to account for IBP. 

In section~\ref{appsubsec:example} we highlight a simple discrepancy in the operator class $H^4 X\mathcal{D}^2$ by fully reconstructing the SM EFT operators and finding operators not present in LM. In section~\ref{appsubsec:ibp}, we explicitly show how the LM analysis over counts the IBP relations in this class. Finally in section~\ref{appsubsec:list}, we give a full list of dim-8 operators with two and three derivatives, highlighting the additional $62$ operators (for $N_f=1$) to be added to the analysis of LM.

\subsection{A quick example: operator class $H^4 X\mathcal{D}^2$}\label{appsubsec:example}

The Hilbert series using our method and that of LM gives:
\begin{align}
\text{Our analysis: }&  \widehat{H}_8 \supset 2H^2H^{\dagger2} W^L \mathcal{D}^2 + H^2H^{\dagger2} B^L \mathcal{D}^2 + \ \text{h.c.} \label{eq:compare_us}\\
\text{LM's analysis: }& \widehat{H}_8 \supset \ \, H^2H^{\dagger2} W^L \mathcal{D}^2 + \ \text{h.c.} \label{eq:compare_LM}
\end{align}
It is straightforward to explicitly construct the operators in the class $H^4 X\mathcal{D}^2$; one finds that the following three operators and their hermitian conjugates\footnote{To be clear, hermitian conjugation includes replacing \(X_{L} \leftrightarrow X_R\) for field strengths, even though in Euclidean space they are not hermitian conjugates in the usual sense.} are all independent:
\begin{eqnarray}
&\left[(D^\mu H)^\dagger (D^\nu H)\right] (H^\dagger H) B_{\mu\nu}^L& \\
&\left[(D^\mu H)^\dagger \t^a (D^\nu H)\right] (H^\dagger H) W_{\mu\nu}^{L,a},\qquad \left[(D^\mu H)^\dagger (D^\nu H)\right] (H^\dagger \t^a H) W_{\mu\nu}^{L,a}& .
\end{eqnarray}
{\it i.e.} there are six operators in this class, which is consistent with our result in eq.~\eqref{eq:compare_us} and table~\ref{tbl:dim8}. The LM analysis does not find the operator $H^2H^{\dagger2} B^L \mathcal{D}^2$ and has a different counting of the operator $H^2H^{\dagger2} W^L \mathcal{D}^2$.

\subsection{LM over counts IBP relations}\label{appsubsec:ibp}

Here we will show how IBPs are over-counted for the operator class $H^2 H^{\dagger2} B^L \mathcal{D}^2$ discussed above. Without imposing IBP, there are two independent operators:
\begin{equation}
\left[(D^\mu H)^\dagger (D^\nu H)\right] (H^\dagger H) B_{\mu\nu}^L, \qquad \left[(H^\dagger D^\mu H)\right]\left[(D^\nu H)^\dagger H\right] B_{\mu\nu}^L .
\end{equation}
With one less power of $\mathcal{D}$, there are also two:
\begin{equation}
A_{1\mu}=\left[H^\dagger (D^\nu H)\right] (H^\dagger H) B_{\mu\nu}^L, \qquad A_{2\mu}=\left[(H^\dagger H)\right]\left[(D^\nu H)^\dagger H\right] B_{\mu\nu}^L .
\end{equation}
In the LM analysis, both $A_{1\mu}$ and $A_{2\mu}$ are assumed to give an independent IBP relation (recall, IBP relations arise from \(0 = \pd^{\m}A_{i\m}\)). Therefore after removing IBP redundancy, it gives the number of operators as $2-2=0$. However, one can check that the IBP relations generated by $A_{1\mu}$ and $A_{2\mu}$ are linearly related, \textit{i.e.} $\pd^\mu A_{1\mu} = -\pd^\mu A_{2\mu}$. This happens because one linear combination of $A_{1\mu}$ and $A_{2\mu}$ can be written as a divergence over a 2-from:
\begin{equation}
A_\mu\equiv A_{1\mu} + A_{2\mu}= \pd^\nu\left[ \frac{1}{2} (H^\dagger H)^2 B_{\mu\nu}^L \right]=\pd^\nu C_{\mu\nu} , \label{eqn:divergence}
\end{equation}
whose total derivative $\pd^\mu A_\mu$ then will be identically zero by symmetry, and hence does not generate an IBP relation. In the Hodge dual language of section~\ref{subsec:HodgeDual}, eq.~\eqref{eqn:divergence} means the dual of $A_\mu$ (which is a 3-form) is the exterior derivative of the dual of $C_{\mu\nu}$ (which is a 2-form):\footnote{Here ``$\sim$'' means we are ignoring any proportional coefficients.}
\begin{equation}
*A \sim \text{d} (*C) ,
\end{equation}
namely that $*A$ is an exact 3-form, whose exterior derivative is identically zero and does not generate an exact 4-form (Hodge dual of IBP operator):
\begin{equation}
*(\pd^\mu A_\mu) \sim \text{d} (*A) \sim \text{d}\left[\text{d} (*C)\right] =0 .
\end{equation}
This is an explicit example where LM counts both non-exact and exact 3-forms. However, exact 3-forms---such as the $*A$ above---do not generate an IBP relation.

\subsection{The list of dim-8 operators with two and three derivatives}\label{appsubsec:list}

In this subsection, we list out all the independent dim-8 operators involving two and three derivatives, highlighting the $62$ operators (for $N_f=1$) not found by the LM analysis.

\subsubsection*{Two Derivatives}%
For classes involving two derivatives, we find an additional 60 operators compared to the LM analysis. We find both larger coefficients in the Hilbert series for some of the operators found in LM, and also operators which were missing entirely. We present the general $N_f$ case and highlight the differences with LM in the case $N_f=1$ in the text. \newline\newline{\small
\underline{Class $H^6\mathcal{D}^2$:}
\newline Self conjugate:
\be&&
2H^{3}H^{\dag\,3}\mathcal{D}^2\ee
\underline{Class $H^4X\mathcal{D}^2$:}
\be&&
H^{2}H^{\dag\,2}B_L\mathcal{D}^2,~2H^{2}H^{\dag\,2}W_L\mathcal{D}^2~~~\text{all + h.c.}
\ee
}The first of these operators was missing from LM; we also find an additional operator of the second type. This class was discussed in the previous subsections. This equates to 4 additional operators of this class. \newline\newline{\small
\underline{Class $H^3\psi^2\mathcal{D}^2$:}
\be&&
6 N_f^2dQHH^{\dag\,2}\mathcal{D}^2,~6 N_f^2eLHH^{\dag\,2}\mathcal{D}^2,~6 N_f^2uQH^{2}H^{\dag}\mathcal{D}^2~~~\text{all + h.c.}
\ee
}Compared with LM we find one additional operator of each type (coefficients 6 vs. coefficients 5). This equates to 6 additional operators of this class.\newline\newline{\small
\underline{Class $H^2X^2\mathcal{D}^2$:}
\newline Self conjugate:
\be&&
HH^{\dag}B_LB_R\mathcal{D}^2,~2HH^{\dag}W_LW_R\mathcal{D}^2,~HH^{\dag}G_LG_R\mathcal{D}^2\ee
and
 \be&&
HH^{\dag}B_L^{2}\mathcal{D}^2,~2HH^{\dag}B_LW_L\mathcal{D}^2,~HH^{\dag}B_RW_L\mathcal{D}^2,~2HH^{\dag}W_L^{2}\mathcal{D}^2,~\nonumber \\&&
HH^{\dag}G_L^{2}\mathcal{D}^2~~~\text{all + h.c.}
\ee
}In LM all these operators were found with coefficient 1; here we find two of them with coefficient 2. This equates to 4 additional operators of this class.\newline\newline {\small
\underline{Class $HX\psi^2\mathcal{D}^2$:}
\be&&
2 N_f^2dQH^{\dag}B_L\mathcal{D}^2,~2 N_f^2eLH^{\dag}B_L\mathcal{D}^2,~2 N_f^2uQHB_L\mathcal{D}^2,~N_f^2dQH^{\dag}B_R\mathcal{D}^2,~\nonumber \\&&
N_f^2eLH^{\dag}B_R\mathcal{D}^2,~N_f^2uQHB_R\mathcal{D}^2,~2 N_f^2dQH^{\dag}W_L\mathcal{D}^2,~2 N_f^2eLH^{\dag}W_L\mathcal{D}^2,~\nonumber \\&&
2 N_f^2uQHW_L\mathcal{D}^2,~N_f^2u^{\dag}Q^{\dag}H^{\dag}W_L\mathcal{D}^2,~N_f^2d^{\dag}Q^{\dag}HW_L\mathcal{D}^2,~N_f^2e^{\dag}L^{\dag}HW_L\mathcal{D}^2,~\nonumber \\&&
2 N_f^2dQH^{\dag}G_L\mathcal{D}^2,~2 N_f^2uQHG_L\mathcal{D}^2,~N_f^2dQH^{\dag}G_R\mathcal{D}^2,~N_f^2uQHG_R\mathcal{D}^2\nonumber \\&&
~~~\text{all + h.c.}
\ee
}All of the operators in this class that we find with coefficient 2 were found with coefficient 1 in LM; all other operators were missing from LM. This equates to 32 additional operators of this class.\newline\newline{\small
\underline{Class $\psi^4\mathcal{D}^2$:}
\newline Self conjugate:
\be&&
(N_f^4+N_f^2)d^2d^{\dag\,2}\mathcal{D}^2,~2 N_f^4dd^{\dag}ee^{\dag}\mathcal{D}^2,~4 N_f^4dd^{\dag}uu^{\dag}\mathcal{D}^2,~2 N_f^4dd^{\dag}LL^{\dag}\mathcal{D}^2,~\nonumber \\&&
4 N_f^4dd^{\dag}QQ^{\dag}\mathcal{D}^2,~\frac{1}{2} \left(N_f^4+N_f^2\right)e^{2}e^{\dag\,2}\mathcal{D}^2,~2 N_f^4ee^{\dag}uu^{\dag}\mathcal{D}^2,~2 N_f^4ee^{\dag}LL^{\dag}\mathcal{D}^2,~\nonumber \\&&
2 N_f^4ee^{\dag}QQ^{\dag}\mathcal{D}^2,~(N_f^4+N_f^2)u^{2}u^{\dag\,2}\mathcal{D}^2,~2 N_f^4uu^{\dag}LL^{\dag}\mathcal{D}^2,~4 N_f^4uu^{\dag}QQ^{\dag}\mathcal{D}^2,~\nonumber \\&&
(N_f^4+N_f^2)L^{2}L^{\dag\,2}\mathcal{D}^2,~4 N_f^4QQ^{\dag}LL^{\dag}\mathcal{D}^2,~2 \left(N_f^4+N_f^2\right)Q^{2}Q^{\dag\,2}\mathcal{D}^2\ee
and
 \be&&
3 N_f^4duQ^{2}\mathcal{D}^2,~2 N_f^4de^{\dag}QL^{\dag}\mathcal{D}^2,~3 N_f^4euQL\mathcal{D}^2~~~\text{all + h.c.}
\ee
}The first of these operators was not found in LM. The last of these operators was found but with coefficient 2. This equates to 8 additional operators.\newline{\small
Baryon number violating terms
 \be&&
2 N_f^4duQ^{\dag}L^{\dag}\mathcal{D}^2,~\frac{1}{2} N_f^3 (3 N_f-1)deu^{2}\mathcal{D}^2,~N_f^4euQ^{\dag\,2}\mathcal{D}^2,~N_f^4Q^{3}L\mathcal{D}^2\nonumber \\&&
~~~\text{all + h.c.}
\ee
}The first of these operators agrees with LM; the last three were not found in their analysis. This equates to 6 additional operators for $N_f=1$. In total we find an additional 14 operators in this class.{\small
\subsubsection*{Three Derivatives}%
}We find one additional type of operator with three derivatives that was not present in the LM analysis, accounting for the remaining `+2' in the discrepancy.\newline\newline{\small
\underline{Class $H^2\psi^2\mathcal{D}^3$:}
\newline Self conjugate:
\be&&
2 N_f^2dd^{\dag}HH^{\dag}\mathcal{D}^3,~2 N_f^2ee^{\dag}HH^{\dag}\mathcal{D}^3,~2 N_f^2uu^{\dag}HH^{\dag}\mathcal{D}^3,~4 N_f^2LL^{\dag}HH^{\dag}\mathcal{D}^3,~\nonumber \\&&
4 N_f^2QQ^{\dag}HH^{\dag}\mathcal{D}^3\ee
and
 \be&&
N_f^2d^{\dag}uH^{2}\mathcal{D}^3~~~\text{+ h.c.}
\ee
}This is the operator (+h.c.) that was not found in LM.

\bibliography{./bibliography}
\bibliographystyle{JHEP}

\end{document}